\documentclass[a4paper,usenames,dvipsnames,11pt]{article}
\pdfoutput=1 % use larger type; default would be 10pt
\usepackage[left=25mm, top=25mm, bottom=25mm, right=25mm]{geometry}  % Set page margins
\usepackage[T1]{fontenc}
\usepackage[utf8]{inputenc} % set input encoding (not needed with XeLaTeX)

\usepackage{graphicx%,wrapfig
} % support the \includegraphics command and options
\usepackage[dvipsnames]{xcolor}
\usepackage{booktabs} % for much better looking tables
\usepackage{colortbl}
\usepackage{array} % for better arrays (eg matrices) in maths
\usepackage{hyperref}
\usepackage{tikz}
\usetikzlibrary{decorations.pathmorphing}
\usetikzlibrary{decorations.markings}
\usepackage{verbatim} % adds environment for commenting out blocks of text & for better verbatim
\usepackage{amsmath}
\usepackage{mathtools}
\usepackage{booktabs}
\usepackage{amssymb}
\newcommand{\Pwr}{Q}

\def\mytitle{A particle physicist lacing their shoes}

\title{\mytitle}
\author{Rodrigo Alonso\\
Institute for Particle Physics Phenomenology, Durham University, \\Durham, DH1 3LE, United Kingdom}

\begin{document}
\begin{titlepage}
\thispagestyle{empty}

\begin{flushright}
IPPP/25/17 % Preprint number added
\end{flushright}

\vspace{15mm}

\begin{center}
{\LARGE\textbf{\mytitle}}
%\maketitle
\renewcommand*{\thefootnote}{\fnsymbol{footnote}}

\vspace{1.1cm}
Rodrigo Alonso\\
Institute for Particle Physics Phenomenology\\ Durham University, DH1 3LE, United Kingdom\\
rodrigo.alonso-de-pablo@durham.ac.uk
\end{center}
\vspace{8mm}

\begin{abstract}
This letter presents the solution to a counting problem, to the best of our knowledge not known in full generality, which can be mapped to both (i) ways to lace a $N$-holes-per-side shoe with $\ell$ shoestrings and  (ii) a sum over the indexes of $2N$-tensors made of symmetric 2-tensors. The coefficients that answer this question are then connected to spin $J=N$ boson exchange amplitudes and perturbative unitarity constraints, deriving relations with Legendre polynomials and sum rules.
\end{abstract}
\end{titlepage}

\section{Lacing your shoes}
Consider one of your shoes with $N$ holes on the left side and another $N$ on the right for the shoestrings to go through, shoestrings which we lace as follows:
\begin{itemize}
\item[i)] The two ends of the shoestring coming out of a hole should go to holes on the other side,
\item[ii)] every hole must have one string running through it,
\item[iii)] after lacing, the two ends of a shoestring are tied together which we take to define a closed string loop. 
\end{itemize}
A pertinent remark  is that there need not be a single shoestring as in your actual shoe but $\ell$ of them, with $1\leq\ell\leq N$.

The question this letter answers is, how many topologically different ways are there to tie your laces? By topological we mean we put on the same class
\begin{itemize}
\item[{\it a})] configurations that differ by reversing the direction of the shoestring through a given hole (i.e. a shoestring from point A on the the opposite side coming in from the top and going out from under to a point B and the configuration with exchanged $A\leftrightarrow B$),
\item[{\it b})] configurations that can be taken to one another by a permutation of the $N$ holes on one side.
\end{itemize}

The first few $N$'s are drawn in eqs.~(\ref{Shoe1}-\ref{Shoe3b}), up and bottom being the two sides of the shoe with holes marked as small circles. For $N=1$ the laces out of a hole have to go to the only other hole but it could be that the string coming out on top on one side goes in from the top or bottom on the other for a total of two. Due to {\it a} above we  group such instances into the same class so in eq.~(\ref{Shoe1}) we count $2$ configurations with $\ell=1$. For $N=2$ and two loops ($\ell=2$), one can have the first hole on the upper side and first on the lower being connected, or the first and second as the first two terms of eq.~(\ref{Shoe2a}) show; given {\it b} the two are equivalent and classed as the first term in eq.~(\ref{Shoe2b}). Analogously for $N=3$ and two shoestrings ($\ell=2$) eq.~(\ref{Shoe3a}) shows two configurations (the last two of the line) grouped in the same class in the second term of eq.~(\ref{Shoe3b})
\begin{align}
N=1\,,\quad&
\raisebox{-5mm}{\begin{tikzpicture}
\draw[very thick,brown] (0,0)	circle (3pt);
\draw[very thick,brown] (0,1)	circle (3pt);
	\draw [thick, rounded corners=3pt] (0,0)--(.5,0.1)--(.5,0.9)--(0,1);
\draw [thick, rounded corners=3pt] (-0.1,0)--(-.5,0.1)--(-.5,0.9)--(-0.1,1);
\end{tikzpicture}}\,\,\,+\,\,
\raisebox{-5mm}{
\begin{tikzpicture}
\draw[very thick,brown] (0,0)	circle (3pt);
\draw[very thick,brown] (0,1)	circle (3pt);
	\draw [thick, rounded corners=3pt] (0,0)--(.5,0.1)--(.5,0.9)--(0.1,1);
\draw [thick, rounded corners=3pt] (-0.1,0)--(-.5,0.1)--(-.5,0.9)--(-0,1);
\end{tikzpicture}
}=\,\,\,2\,\,\,\raisebox{-5mm}{
\begin{tikzpicture}
\draw[very thick,brown] (0,0)	circle (1pt);
\draw[very thick,brown] (0,1)	circle (1pt);
	\draw [thick, rounded corners=3pt] (0,0)--(.5,0.1)--(.5,0.9)--(0,1);
\draw [thick, rounded corners=3pt] (0,0)--(-.5,0.1)--(-.5,0.9)--(-0,1);
\end{tikzpicture}}\,\,\,.\label{Shoe1}
\\
N=2\,,\quad&
\raisebox{-5mm}{\begin{tikzpicture}
\draw[very thick,brown] (0,0)	circle (1pt);
\draw[very thick,brown] (0,1)	circle (1pt);
	\draw [thick, rounded corners=3pt] (0,0)--(.5,0.1)--(.5,0.9)--(0,1);
\draw [thick, rounded corners=3pt] (0,0)--(-.5,0.1)--(-.5,0.9)--(0,1);
\end{tikzpicture}
\begin{tikzpicture}
\draw[very thick,brown] (0,0)	circle (1pt);
\draw[very thick,brown] (0,1)	circle (1pt);
	\draw [thick, rounded corners=3pt] (0,0)--(.5,0.1)--(.5,0.9)--(0,1);
\draw [thick, rounded corners=3pt] (0,0)--(-.5,0.1)--(-.5,0.9)--(-0,1);
\end{tikzpicture}
}+
\raisebox{-5mm}{\begin{tikzpicture}
\draw[very thick,brown] (0,0)	circle (1pt);
\draw[very thick,brown] (0,1)	circle (1pt);
\draw[very thick,brown] (1,0)	circle (1pt);
\draw[very thick,brown] (1,1)	circle (1pt);
	\draw [thick, rounded corners=3pt] (0,0)--(0.75,0.2)--(1,1);
\draw [thick, rounded corners=3pt] (0,0)--(.2,.75)--(1,1);
	\draw [thick, rounded corners=3pt] (0,0)--(0.75,0.2)--(1,1);
	\draw [thick, rounded corners=3pt] (0,1)--(0.25,0.25)--(1,0);
	\draw [thick, rounded corners=3pt] (0,1)--(.75,.75)--(1,0);
\end{tikzpicture}
}+\raisebox{-5mm}{
\begin{tikzpicture}
\draw[very thick,brown] (0,0)	circle (1pt);
\draw[very thick,brown] (0,1)	circle (1pt);
\draw[very thick,brown] (1,0)	circle (1pt);
\draw[very thick,brown] (1,1)	circle (1pt);
	\draw [thick, rounded corners=3pt] (0,0)--(1,1);
	\draw [thick, rounded corners=3pt] (0,0)--(-.5,0.1)--(-.5,0.9)--(0,1);
	\draw [thick, rounded corners=3pt] (0,1)--(1,0);
	\draw [thick, rounded corners=3pt] (1,0)--(1.5,0.1)--(1.5,0.9)--(1,1);
\end{tikzpicture}\label{Shoe2a}
}+\cdots\\
&=4\left(2\,\,\raisebox{-5mm}{\begin{tikzpicture}
\draw[very thick,brown] (0,0)	circle (1pt);
\draw[very thick,brown] (0,1)	circle (1pt);
	\draw [thick, rounded corners=3pt] (0,0)--(.5,0.1)--(.5,0.9)--(0,1);
\draw [thick, rounded corners=3pt] (0,0)--(-.5,0.1)--(-.5,0.9)--(0,1);
\end{tikzpicture}
\begin{tikzpicture}
\draw[very thick,brown] (0,0)	circle (1pt);
\draw[very thick,brown] (0,1)	circle (1pt);
	\draw [thick, rounded corners=3pt] (0,0)--(.5,0.1)--(.5,0.9)--(0,1);
\draw [thick, rounded corners=3pt] (0,0)--(-.5,0.1)--(-.5,0.9)--(-0,1);
\end{tikzpicture}
}+4\,\,\raisebox{-5mm}{
\begin{tikzpicture}
\draw[very thick,brown] (0,0)	circle (1pt);
\draw[very thick,brown] (0,1)	circle (1pt);
\draw[very thick,brown] (1,0)	circle (1pt);
\draw[very thick,brown] (1,1)	circle (1pt);
	\draw [thick, rounded corners=3pt] (0,0)--(1,1);
	\draw [thick, rounded corners=3pt] (0,0)--(-.5,0.1)--(-.5,0.9)--(0,1);
	\draw [thick, rounded corners=3pt] (0,1)--(1,0);
	\draw [thick, rounded corners=3pt] (1,0)--(1.5,0.1)--(1.5,0.9)--(1,1);
\end{tikzpicture}\label{Shoe2b}
}\right)\,\,.\\
N=3\,,\quad&
\raisebox{-5mm}{\begin{tikzpicture}
\draw[very thick,brown] (0,0)	circle (1pt);
\draw[very thick,brown] (0,1)	circle (1pt);
	\draw [thick, rounded corners=3pt] (0,0)--(.5,0.1)--(.5,0.9)--(0,1);
\draw [thick, rounded corners=3pt] (0,0)--(-.5,0.1)--(-.5,0.9)--(0,1);
\end{tikzpicture}
\begin{tikzpicture}
\draw[very thick,brown] (0,0)	circle (1pt);
\draw[very thick,brown] (0,1)	circle (1pt);
	\draw [thick, rounded corners=3pt] (0,0)--(.5,0.1)--(.5,0.9)--(0,1);
\draw [thick, rounded corners=3pt] (0,0)--(-.5,0.1)--(-.5,0.9)--(-0,1);
\end{tikzpicture}
\begin{tikzpicture}
\draw[very thick,brown] (0,0)	circle (1pt);
\draw[very thick,brown] (0,1)	circle (1pt);
	\draw [thick, rounded corners=3pt] (0,0)--(.5,0.1)--(.5,0.9)--(0,1);
\draw [thick, rounded corners=3pt] (0,0)--(-.5,0.1)--(-.5,0.9)--(-0,1);
\end{tikzpicture}
}
+\raisebox{-5mm}{
\begin{tikzpicture}
\draw[very thick,brown] (0,0)	circle (1pt);
\draw[very thick,brown] (0,1)	circle (1pt);
\draw[very thick,brown] (1,0)	circle (1pt);
\draw[very thick,brown] (1,1)	circle (1pt);
	\draw [thick, rounded corners=3pt] (0,0)--(1,1);
	\draw [thick, rounded corners=3pt] (0,0)--(-.5,0.1)--(-.5,0.9)--(0,1);
	\draw [thick, rounded corners=3pt] (0,1)--(1,0);
	\draw [thick, rounded corners=3pt] (1,0)--(1.5,0.1)--(1.5,0.9)--(1,1);
\end{tikzpicture}
\begin{tikzpicture}
\draw[very thick,brown] (0,0)	circle (1pt);
\draw[very thick,brown] (0,1)	circle (1pt);
	\draw [thick, rounded corners=3pt] (0,0)--(.5,0.1)--(.5,0.9)--(0,1);
\draw [thick, rounded corners=3pt] (0,0)--(-.5,0.1)--(-.5,0.9)--(-0,1);
\end{tikzpicture}
}
+\raisebox{-5mm}{
\begin{tikzpicture}
\draw[very thick,brown] (0,0)	circle (1pt);
\draw[very thick,brown] (0,1)	circle (1pt);
\draw[very thick,brown] (1,0)	circle (1pt);
\draw[very thick,brown] (1,1)	circle (1pt);
\draw[very thick,brown] (2,0)	circle (1pt);
\draw[very thick,brown] (2,1)	circle (1pt);
	\draw [thick, rounded corners=3pt] (0,0)--(2,1);
	\draw [thick, rounded corners=3pt] (0,0)--(-.5,0.1)--(-.5,0.9)--(0,1);
	\draw [thick, rounded corners=3pt] (0,1)--(2,0);
	\draw [thick, rounded corners=3pt] (2,0)--(2.5,0.1)--(2.5,0.9)--(2,1);
	\draw [thick, rounded corners=3pt] (1,0)--(1.5,0.1)--(1.5,0.9)--(1,1);
	\draw [thick, rounded corners=3pt] (1,0)--(.5,0.1)--(.5,0.9)--(1,1);
\end{tikzpicture}
}+\dots\label{Shoe3a}
\\\label{Shoe3b}
&=6!!\left(
\raisebox{-5mm}{\begin{tikzpicture}
\draw[very thick,brown] (0,0)	circle (1pt);
\draw[very thick,brown] (0,1)	circle (1pt);
	\draw [thick, rounded corners=3pt] (0,0)--(.5,0.1)--(.5,0.9)--(0,1);
\draw [thick, rounded corners=3pt] (0,0)--(-.5,0.1)--(-.5,0.9)--(0,1);
\end{tikzpicture}
\begin{tikzpicture}
\draw[very thick,brown] (0,0)	circle (1pt);
\draw[very thick,brown] (0,1)	circle (1pt);
	\draw [thick, rounded corners=3pt] (0,0)--(.5,0.1)--(.5,0.9)--(0,1);
\draw [thick, rounded corners=3pt] (0,0)--(-.5,0.1)--(-.5,0.9)--(-0,1);
\end{tikzpicture}
\begin{tikzpicture}
\draw[very thick,brown] (0,0)	circle (1pt);
\draw[very thick,brown] (0,1)	circle (1pt);
	\draw [thick, rounded corners=3pt] (0,0)--(.5,0.1)--(.5,0.9)--(0,1);
\draw [thick, rounded corners=3pt] (0,0)--(-.5,0.1)--(-.5,0.9)--(-0,1);
\end{tikzpicture}
}
+6\,\raisebox{-5mm}{
\begin{tikzpicture}
\draw[very thick,brown] (0,0)	circle (1pt);
\draw[very thick,brown] (0,1)	circle (1pt);
	\draw [thick, rounded corners=3pt] (0,0)--(.5,0.1)--(.5,0.9)--(0,1);
\draw [thick, rounded corners=3pt] (0,0)--(-.5,0.1)--(-.5,0.9)--(-0,1);
\end{tikzpicture}
\begin{tikzpicture}
\draw[very thick,brown] (0,0)	circle (1pt);
\draw[very thick,brown] (0,1)	circle (1pt);
\draw[very thick,brown] (1,0)	circle (1pt);
\draw[very thick,brown] (1,1)	circle (1pt);
	\draw [thick, rounded corners=3pt] (0,0)--(1,1);
	\draw [thick, rounded corners=3pt] (0,0)--(-.5,0.1)--(-.5,0.9)--(0,1);
	\draw [thick, rounded corners=3pt] (0,1)--(1,0);
	\draw [thick, rounded corners=3pt] (1,0)--(1.5,0.1)--(1.5,0.9)--(1,1);
\end{tikzpicture}
}
+8\,\raisebox{-5mm}{
\begin{tikzpicture}
\draw[very thick,brown] (0,0)	circle (1pt);
\draw[very thick,brown] (0,1)	circle (1pt);
\draw[very thick,brown] (1,0)	circle (1pt);
\draw[very thick,brown] (1,1)	circle (1pt);
\draw[very thick,brown] (2,0)	circle (1pt);
\draw[very thick,brown] (2,1)	circle (1pt);
	\draw [thick, rounded corners=3pt] (0,0)--(1,1);
	\draw [thick, rounded corners=3pt] (0,0)--(-.5,0.1)--(-.5,0.9)--(0,1);
	\draw [thick, rounded corners=3pt] (0,1)--(1,0);
	\draw [thick, rounded corners=3pt] (2,0)--(2.5,0.1)--(2.5,0.9)--(2,1);
	\draw [thick, rounded corners=3pt] (1,0)--(2,1);
	\draw [thick, rounded corners=3pt] (1,1)--(2,0);
\end{tikzpicture}
}
\right)\,.
\end{align}

Let us then define $\hat c_{p_1\dots p_\ell}^{(N)}$ with ordered indexes $p_i\leq p_{i+1}$ as the symbol counting configurations with  $\ell$ shoestrings, the number i'th of them going through $2p_i$ holes, noting that $p_i$ are constrained by $\sum p_i=N$. Given our rules for lacing i-iii and the equivalence relations {\it a-b}, these indexes provide all input needed to characterise the different classes.
We have found thus far by explicit computation
\begin{align}
\hat c_{1}^{(1)}&=2\,,  & 
\hat c_{11}^{(2)}&=4!!\,, &
\hat c_{2}^{(2)}&=4!!\times2\,, \\
\hat c_{111}^{(3)}&=6!!\,, &
\hat c_{12}^{(3)}&=6!!\times 6\,, &
\hat c_{3}^{(3)}&=6!!\times 8\,.
\end{align}
For convenience we define
\begin{align}
 c_{p_1\dots p_\ell}^{(N)}=\frac{1}{(2N)!!}\hat c_{p_1\dots p_\ell}^{(N)}\,.\label{LacesEq}
\end{align}
 The main result of this letter is an explicit formula for $c^{(N)}_{p_1\dots p_{\ell}}$, to the best of our knowledge not previously known. In this regard let us note that the first few elements of $c^{(N)}_{p_1\dots p_{\ell}}$ do not return any known sequence in the Online Encyclopedia of Integer Sequences \cite{OEIS}. Our best efforts at finding this same problem in graph theory through ref.~\cite{Diestel} led to the identification of the type of diagrams here considered as order $2N$ degree $2$  (for all vertexes) bipartite graphs which we characterise by the number of cycles $\ell$ and their length $2p_i$ yet we could find no formulation of the problem here considered.

\section{Tensor formulation}
We can formulate a related problem with tensors, consider a $2N$-tensor $T^{(L)}$ 
\begin{align}
T^{(L)}_{i_1\dots i_{2N}}\equiv L_{(i_1 i_2}\times L_{i_3i_4}\times\cdots\times L_{i_{2N-1}i_{2N})}\,,
\end{align}
where the 2-tensor $L$ is symmetric ($L_{ij}=L_{ji}$), the ellipsis in $i_1\dots i_{2N}$ stands for all the indexes in between, and the parenthesis stands for symmetrisation in the indexes which can be built recursively as
\begin{align}
T^{(L)}_{(i_1\dots i_{2N})}=\sum_{k=1}^{2N-1}T_{(i_1\dots i_{2N}\dots i_{2N-2})}L_{i_{2N-1}i_{k}}\,,
\end{align}
i.e. replace the index in the $k$'th position by $i_{2N}$ including the index $2N-1$ in $L$. The starting point of the iteration is $T_{ij}=L_{ij}$ and one finds $(2N-1)!!$ terms in the sum above at order $N$. 

The problem then is to find the coefficients
\begin{align}
\sum_{i_1\dots i_{2N}}T_{i_1\dots i_{2N}}^{(L)}T^{(R)}_{i_1\dots i_{2N}}=\sum_{\ell=1}^{N}\sum_{p_1\leq \dots \leq p_\ell}d_{p_1\dots p_\ell}^{(N)} \prod_k \mbox{Tr}((LR)^{p_k})\,,
\end{align}
where $\sum p_k=N$ and $p_k$ are the same characterising lenghts of eq.~(\ref{LacesEq}) since 
\begin{align}
d_{p_1\dots p_\ell}^{(N)}=(2N-1)!! \,c_{p_1\dots p_\ell}^{(N)}\,.
\end{align}
only now $p_k$ count the powers of $LR$ inside the traces.

\section{Explicit formula}\label{Glory}
For the explicit form of $c$ let us split the $p_i$ indexes into the first $\ell-\ell'$ of them equal to $1$, and the remaining $\ell'$ as
\begin{align}
c_{p_1\cdots p_\ell}=c_{1\dots 1 p'_1 \dots p'_{\ell'}}\,,
\end{align}
that is $p'_i$ are the subset of $p_j$ such that $p_j\geq 2$.

The solution then reads, with $\mathbf{p}'$ the array of $p'$s ($\mathbf{p}'=(p_1',p_2',\dots,p_{\ell'}')$)
\begin{align}
c_{p_1\cdots p_\ell}=c_{1\dots 1 p'_1\dots p'_{\ell'}}
=\sum_{ \sigma(\mathbf{p'})=\mathbf{q}}\sum_{m_1,\dots, m_{\ell'}}\prod_{k=1}^{\ell'}\frac{\left(2\left[N-1-m_k-\sum_j^{k-1}(m_j+q_j)\right]\right)!!}{\left(2\left[N-\sum_j^{k}(m_j+q_j)\right]\right)!!}\,,
\end{align}
%\begin{align}
%\sum_{m_1,\dots, m_{\ell'}}\equiv\sum_{m_1=1}^{N-q_1}\sum_{m_2}^{N-m_1-q_1-q_2}\cdots\sum_{m_{\ell'}}^{N-q_{\ell'}-\sum^{\ell'-1}(m_k+q'_k)}
%\end{align}
where the outer sum is over permutations $\sigma(\mathbf{p}')$ and is to be understood as follows for an explicit example of $\mathbf{p}'=(2,2,3)$ (i.e. $p_1'=2$, $p_2'=2$, $p_3'=3$)
\begin{align}
\sum_{ \sigma(223)=\mathbf{q}}F(\mathbf{q})=F\left(
\mathbf{q}=(2,2,3)
\right)+F\left(
\mathbf{q}=(2,3,2)
\right)+F\left(
\mathbf{q}=(3,2,2)
\right)\,,
\end{align}
whereas $\mathbf{p}'=(2,3,4)$ instead would give rise to six terms. Lastly the second summation runs from $m_k=0$ until the boundary defined by positive arguments for the double factorials in the numerators, i.e
\begin{align}
N-\sum_{j=1}^{k}(m_j+q_j)&\geq 0 & \forall k\,.
\end{align}
%For example in the same configuration $223$
%\begin{align}
%\frac{(2(N-m_1))!!}{(2(N-m_1-1))!!}
%\frac{(2(N-m_1-m_2-1))!!}{(2(N-m_1-m_2-2))!!}
%\frac{(2(N-m_1-m_2-m_3-2))!!}{(2(N-m_3-m_2-m_1-4))!!}
%\end{align}
The only element this formula does not apply to in an straightforward way is $c^{(N)}_{1\dots 1}$, i.e. the element with $N=\ell$, this however is simply $c^{(N)}_{1\dots 1}=1$ for all $N$. 
\section{Applications in Physics}
Here are two instances in which the solution to the counting problem above has applications in theoretical particle physics. The emphasis is on the appearance of the constants $c^{(N)}_{p_1\dots p_\ell}$, the theory background is scattering theory, see e.g.~\cite{Eden:1966dnq} for a standard reference and for more recent work pertinent to sec.~4.1,~\cite{Arkani-Hamed:2017jhn}. 

\subsection{Exchange of spin $J=N$ particle}

The polarisation of a spin $J$ and mass $M$ particle in $4$ spacetime dimensions can be specified by  symmetrised indexes $i_1,i_2\dots, i_{2J}$ with $i_j=1,2$ so that the number of possibilities for given $J$ is $2J+1$. The 3 point amplitude for the emission of such a particle (represented by a wavy line) with 4-momentum $q$ off of a complex scalar (i.e. spin$=0$) massless particle (dashed line) with  4-momenta $p$ before and $p'$ after the emission reads
\begin{align}
&\raisebox{-10mm}{
\begin{tikzpicture}
\draw [thick,dashed,->] (-1.5,1) node [anchor=north] {$p$} --(-.75,0.75);
\draw [thick,dashed] (-.75,0.75) -- (0,0.5);
\draw [thick,dashed] (1.5,1) node [anchor=north] {$p'$} -- (0.75,0.75);
\draw [thick,dashed,->] (0,0.5)--(0.75,0.75);
\draw [very thick,decorate,decoration=snake] (0,0.5)  --(0,-1)node [anchor=west] {$q\,,\,J$};
\draw  (-1.5,-.5) node {time $\rightarrow$};
\end{tikzpicture}
}&
-i\mathcal A^{(3)}_{i_1\dots i_{2J}}&=ig\frac{MT_{i_1\dots i_{2J}}^{(A)}}{(2J-1)!!} \,,
\end{align}
where $g$ is a coupling constant, $A$ is a matrix that depends on the kinematics and the polarisation as $A_{ij}(p+p',q)$, the momenta satify\footnote{The square of a 4-momentum $p$ is the Minkowski scalar product with itself, this product being $p\cdot q=p^0q^0-q^1p^1-q^2p^2-q^3p^3$} $p=q+p'$, $p^2=(p')^2=0$, $q^2=M^2$ and $A_{ij}=A_{ji}$ as can be shown using the kinematic relations while we normalise $A$ such that Tr$(A\epsilon A \epsilon^T)=2(p+p')^2/M^2$ with $\epsilon$ the Levi-Civita in 2 dimensions and $\epsilon^T$ its transpose.
The exchange of the $J$-particle gives a four point amplitude whose leading singular term can be written as a pole in $q^2$ times the on-shell three point amplitudes 
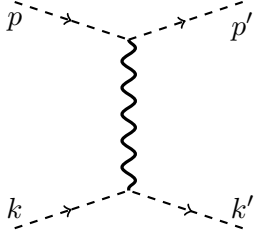
\begin{align}
&\raisebox{-12mm}{
\begin{tikzpicture}
\draw [thick,dashed,->] (-1.5,1) node [anchor=north] {$p$} --(-.75,0.75);
\draw [thick,dashed] (-.75,0.75) -- (0,0.5);
\draw [thick,dashed] (1.5,1) node [anchor=north] {$p'$} -- (0.75,0.75);
\draw [thick,dashed,->] (0,0.5)--(0.75,0.75);
\draw [very thick,decorate,decoration=snake] (0,0.5) --(0,-1.5);
\draw [thick,dashed,->] (-1.5,-2)node [anchor=south] {$k$} -- (-0.75,-1.75);
\draw [thick,dashed] (-0.75,-1.75) -- (0,-1.5);
\draw [thick,dashed,-<] (1.5,-2)node [anchor=south] {$k'$} -- (0.75,-1.75);
\draw [thick,dashed] (0.75,-1.75) -- (0,-1.5);
\end{tikzpicture}
} & \mathcal A^{(4)}&=\frac{gg'M^2}{q^2-M^2}\sum_{i_1\dots i_{2J}}\frac{T_{i_1\dots i_{2J}}^{(A)}}{(2J-1)!!}\frac{T_{i_1\dots i_{2J}}^{(\tilde A)}}{(2J-1)!!}\,,
\end{align}
where $\tilde A=\epsilon A(k+k',-q)\epsilon^T$. The solution to the problem found in sec.~\ref{Glory} will give the coefficients of the product of traces of $A\tilde A$ which itself can be written as polynomials $\hat P_n(x)$ with $x\equiv-(p+p')(k+k')/M^2$ using results of \cite{Alonso:2019ptb} and finally the sum over products of $\hat P$-polynomials yields the Legendre polynomial of degree $J$, $P_J$, as follows
\begin{align}
\mathcal A^{(4)}=&\frac{gg'M^2}{q^2-M^2}\frac{1}{(2J-1)!!}\sum_{\ell=1}^{N}\sum_{p_1\leq \dots \leq p_\ell} c_{p_1\dots p_\ell}^{(J)} \prod_{k=1}^\ell [\textrm{Tr}(A\tilde A)^{p_k}]\\=&\frac{gg'M^2}{q^2-M^2}\frac{1}{(2J-1)!!}\sum_{\ell}\sum_{p_1\leq \dots \leq p_\ell} c_{p_1\dots p_\ell}^{(J)} \prod_k \hat P_{p_k}(x)=\frac{gg'M^2}{q^2-M^2}\frac{2J!!}{(2J-1)!!}P_J(x)\,,\label{Legendre}
\end{align}
with
\begin{align}
\hat P_{n}(x)\equiv 2\sum_{m=0}^{\lfloor n/2\rfloor} \binom{n}{2m} x^{n-2m}(x^2-1)^m\,.
\end{align}
This gives the connection between $c^{(J)}_{p_1\dots p_\ell}$ and Legendre polynomials which can be made explicit as
\begin{align}
\sum_{\ell}\sum_{p_1\leq \dots \leq p_\ell} c^{(J)}_{p_1\dots p_\ell} \prod_k \left(2\sum_{m_k} \binom{p_k}{2m_k} x^{p_k-2m_k}(x^2-1)^{m_k}\right)=
\frac{d^J}{dx^J}(x^2-1)^J\,.
\end{align}

Taking the Fourier transform in 4 dimensions of eq.~(\ref{Legendre}) with $p+p'=k+k'=(M,0,0,0)$ from $q$-space to spacetime $x=(x^0,x^1,x^2,x^3)$ and integrating over time ($x^0$) one obtains the potential generated by the spin-$J$ particle field between a static source at the origin and another at a spatial distance $r=\sqrt{(x^1)^2+(x^2)^2+(x^3)^2}$. This is a potential of the Yukawa form $V\propto- P_J(-1)e^{-Mr}/r=(-1)^{J+1}e^{-Mr}/r$ and leads to the repulsion (attraction) of same charge sources, i.e. sign$(g)=$sign$(g')$, for  odd (even) $J$. This is indeed the known behaviour of gravity (mediated by the graviton with $J=2,M=0$), electromagnetism (mediated by the photon with $J=1,M=0$) and the original idea of Yukawa, pions ($J=0$) as mediators of nuclear interactions. The same amplitude in eq.~(\ref{Legendre}), when changing which particles are incoming and outgoing as $p'\to-p'$, $k\to -k$ and setting $g=g'$, gives scalar-antiscalar annihilation into spin-$J$ boson which on shell returns the partial decay width and hence needs to have the same sign for all $J$~\cite{Alonso:2019ptb}. This same sign is in agreement with eq.~(\ref{Legendre}) since now the kinematics dictate $x=1$ and $P_J(1)=1$.
%%%%%%%%%%%%%%%%%%%%%%%%%%%%%%%%%%%%%%%%%%%%%%%%%%%
\subsection{Unitarity of the $S$-matrix for a $O(n)$ scalar potential}
Consider a $O(n)$ symmetric theory of a scalar with $n$ components $\phi_{i}$ in $d$ spacetime dimensions with an interaction term in the Lagrangian density $\mathcal L$ and $2\Pwr$-point contact-interaction amplitude
\begin{align}
\mathcal L_\Pwr&=-\frac{a_\Pwr}{(2\Pwr)!!}(\phi^2)^\Pwr\,,& -i\mathcal A^{(2\Pwr)}_{i_1\dots i_{2\Pwr}}=&-ia_\Pwr T^{(\delta)}_{i_1\dots i_{2\Pwr}}\,,
\end{align}
where $\delta$ stands for the Kronecker delta and $\phi^2=\sum_i \phi_i^2$.
The amplitude $\mathcal A^{(2\Pwr)}$ gives the perturbative estimate of the nontrivial piece of the scattering matrix $S$, schematically $S=1-i\mathcal{A}$.
The $S$ matrix is a unitary operator, i.e. $SS^\dagger=S^\dagger S=1$, only the indexes of this matrix run over Fock space (i.e. multiparticle states specified by their momenta). Unitarity for the scattering of $\Pwr'\to \Pwr'$ particles reads
\begin{align}
2\textrm{Im}(\mathcal A_{\Pwr'\to \Pwr'})+\int d\Pi_{\Pwr'} |\mathcal{A}_{\Pwr'\to \Pwr'}|^2+\sum_{X\neq \Pwr}\int d\Pi_X|\mathcal{A}_{\Pwr'\to X}|^2=0\,,
\end{align}
where $d\Pi_b$ stands for the phase space measure of $b$ particles. If we apply this relation to an interaction Lagrangian $\mathcal L_{\Pwr'}+\mathcal L_{\Pwr}$ we can write the above equation at tree level with $X=\Pwr$ as,
\begin{align}
2\textrm{Im}(a_{\Pwr'})\,T^{(\delta)}_{\mathbf{i,j}}+\frac{V_{\Pwr'}}{\Pwr'!}|a_{\Pwr'}|^2\sum_{\mathbf{k}}T_{\mathbf{i,k}}^{(\delta)}T_{\mathbf{j,k}}^{(\delta)}+\frac{V_{2\Pwr-\Pwr'}}{(2\Pwr-\Pwr')!}|a_{\Pwr}|^2\sum_{\mathbf{q}}T_{\mathbf{i,q}}^{(\delta)}T_{\mathbf{j,q}}^{(\delta)}=0\,,
\end{align}
with $\mathbf{i}$ ($\mathbf{j}$) the array of $\Pwr'$ indexes labelling incoming (outgoing) particles $\mathbf{i}=(i_1\dots i_{\Pwr'})$ ($\mathbf{j}=(j_1\dots j_{\Pwr'})$), $\mathbf{k}$ ($\mathbf{q}$) an array of $\Pwr'$ ($2\Pwr-\Pwr'$) internal particle indexes we sum over, and $V_b/b!$ the phase space volume for $b$ undistinguishable particles. The resulting tensors need not be the same as the original, yet a simple relation is obtained if we set $\mathbf{i}=\mathbf{j}$ and sum to find
\begin{align}
2C_{\Pwr'}(n)\,\textrm{Im}(a_{\Pwr'})+\frac{(2\Pwr'-1)!!V_{\Pwr'}C_{\Pwr'}(n)}{\Pwr'!}|a_{\Pwr'}|^2+\frac{V_{2\Pwr-\Pwr'}(2\Pwr-1)!!C_{\Pwr}(n)}{(2\Pwr-\Pwr')!}|a_{\Pwr}|^2=0\,,\label{Argand}
\end{align}
where the polynomial $C_{\Pwr}$ is defined as
\begin{align}
C_{\Pwr}(n)\equiv\sum_{i_1\dots i_{2\Pwr}}\frac{T_{i_1\dots i_{2\Pwr}}T_{i_1\dots i_{2\Pwr}}}{(2\Pwr-1)!!}=&\sum_\ell \sum_{p_1\leq \dots \leq p_\ell}c_{p_1\dots p_\ell}^{{(\Pwr)}}\prod_{k=1}^{\ell} n=\sum_\ell n^\ell\sum_{p_1\leq \dots \leq p_\ell}c_{p_1\dots p_\ell}^{(\Pwr)} \,, \end{align}
and we find
\begin{align}
C_\Pwr(n)=&\frac{(n+2(\Pwr-1))!!}{(n-2)!!}\,.
\end{align}

The constraint in eq.~(\ref{Argand}) in the Argand plane of ($x\equiv$Re$(a_{\Pwr'})$, $y\equiv$ Im$(a_{\Pwr'})$) reads  $ (y+R)^2+x^2=R^2(1-\kappa)$, so allowed values for $(x,y)$ lie within a circle centered around $(0,-R)$. For consistency $\kappa$, which is defined positive, should satisfy $\kappa\leq 1$ else the circle shrinks to nothing. We can collect both constraints on the parameters of the theory as
\begin{align}
\frac{({\Pwr'})!}{(2{\Pwr'}-1)!!V_{{\Pwr'}}}&\geq|\textrm{Re}(a_{\Pwr'})|\,, & \frac{(2\Pwr-\Pwr')!({\Pwr'})!C_{\Pwr'}(n) }{V_{{\Pwr'}}(2{\Pwr'}-1)!!V_{2{\Pwr}-{\Pwr'}}(2{\Pwr}-1)!!C_{\Pwr}(n)}\geq &|a_{\Pwr}|^2\,,\label{BoundsUP}
\end{align}
where we note that considering instead $\Pwr\to \Pwr$ scattering we would obtain another two constrains with $\Pwr \leftrightarrow \Pwr '$.

In natural units both the phase space and the coefficients have mass dimension; making the scales explicit $a_N\equiv 1/(\Lambda)^{N(d/2-1)-d/2}$, and $E$ being the CM energy of the multiparticle collision we have
\begin{align}
\frac{(2{\Pwr}-{\Pwr'})!{\Pwr'}!C_{\Pwr'}(n) }{\hat V_{{\Pwr'}}(2{\Pwr'}-1)!!\hat V_{2{\Pwr}-{\Pwr'}}(2{\Pwr}-1)!!C_{\Pwr}(n)}\Lambda^{2{\Pwr}(d-2)-2d}\geq & E^{2{\Pwr}(d-2)-2d}\,,\label{Nonreno}
\end{align}
where the explicit phase-space volume can be found in~\cite{Gehrmann-DeRidder:2003pne},
\begin{align}
 \hat V_m\equiv\frac{ V_m}{ s^{m(d/2-1)-d/2}}=\frac{2\pi}{(4\pi)^{(m-1)d/2}}\frac{(\Gamma[d/2-1])^m}{\Gamma[(m-1)(d/2-1)]\Gamma[m(d/2-1)]}\,.
\end{align}

A relevant remark that follows eq.~(\ref{Nonreno}) is that, if one can  increase the energy $E$ arbitrarily and the exponent $2\Pwr(d-2)-2d$ is positive (a case referred to in the particle physics literature as non-renormalisable) this inequality will stop being respected at some high energy, marking the limit of validity of the theory. The prominent example in nature of such theories is Fermi's interaction with a coupling $a_\Pwr\to G_F\sim 1/v^2$ whose limit of validity is $E\sim v\sim 246$GeV where indeed the electroweak theory of the Standard Model takes over. This result, which we note in Fermi theory involves fermions not scalar fields, one can obtain directly from an analysis of the scaling with energy of amplitudes; the novelty in the present result is that the bounds in eq.~(\ref{BoundsUP}) are computed explicitly including combinatorial factors and, at first order in perturbation in the present scalar theory, are exact and ultimately made possible by our explicit formula for $c^{(N)}_{p_1\dots p_\ell}$.

%%%%%%%%%%%%%%%%%%%%%%
\section*{Appendix}
One has the sum rules
\begin{align}
\sum_\ell \sum_{p_1\leq \dots \leq p_\ell}c_{p_1\dots p_\ell}^{(N)}&=(2N-1)!! \,,
&\sum_{p_1\leq \dots \leq p_\ell}c_{p_1\dots p_\ell}^{(N)}=&\left(\frac{1}{\ell!}\left(\frac{d}{dn}\right)^{\ell} \frac{(n+2(N-1))!!}{(n-2)!!}\right)_{n=0}\,.
\end{align}
For  $N=1$ one has $c_1^{(1)}=1$ and the first few subsequent sets of coefficients are, with the same background color for consecutive entries with the same number of loops $\ell$, 
\begin{center}
$N=2$\,\,\,\begin{tabular}{|>{\columncolor[gray]{0.8}}c|c|}\hline
$c_{11}^{(2)}$& $c_{2}^{(2)}$\\ \hline \hline
1& 2\\\hline
\end{tabular}\qquad$N=3$\,\,\,\begin{tabular}{|>{\columncolor[gray]{0.8}}c|c|>{\columncolor[gray]{0.8}}c|}\hline
$c_{111}^{(3)}$& $c_{12}^{(3)}$ & $c_3^{(3)}$\\ \hline \hline
1& 6 &8\\\hline
\end{tabular}\qquad
$N=4$\,\,\,\begin{tabular}{|>{\columncolor[gray]{0.8}}c|c|>{\columncolor[gray]{0.8}}c|>{\columncolor[gray]{0.8}}c|c|}
\hline
$c_{1111}^{(4)}$&$c_{112}^{(4)}$&$c_{22}^{(4)}$&$c_{13}^{(4)}$&$c_{4}^{(4)}$\\\hline\hline
1& 12& 12&32& 48\\
\hline
\end{tabular}
\end{center}

\begin{center}
$N=5$\,\,\,\begin{tabular}{|>{\columncolor[gray]{0.8}}c|c|>{\columncolor[gray]{0.8}}c|>{\columncolor[gray]{0.8}}c|c|c|>{\columncolor[gray]{0.8}}c|}
\hline
$c_{11111}^{(5)}$&$c_{1112}^{(5)}$&$c_{122}^{(5)}$&$c_{113}^{(5)}$&$c_{23}^{(5)}$&$c_{14}^{(5)}$&$c_5^{(5)}$\\\hline\hline
1& 20&60& 80& 160&240&384\\
\hline
\end{tabular}
\end{center}

\begin{center}
$N=6$\,\,\,\begin{tabular}{|>{\columncolor[gray]{0.8}}c|c|>{\columncolor[gray]{0.8}}c|>{\columncolor[gray]{0.8}}c|c|c|c|>{\columncolor[gray]{0.8}}c|>{\columncolor[gray]{0.8}}c|>{\columncolor[gray]{0.8}}c|c|}
\hline
$c_{111111}^{(6)}$&$c_{11112}^{(6)}$&$c_{1113}^{(6)}$&$c_{1122}^{(6)}$&$c_{222}^{(6)}$&$c_{114}^{(6)}$&$c_{123}^{(6)}$&$c_{33}^{(6)}$&$c_{24}^{(6)}$&$c_{15}^{(6)}$&$c_6^{(6)}$\\\hline\hline
1& 30& 160&180&120& 720&960&640&1440&2304&3840\\
\hline
\end{tabular}\\[2mm]
\end{center}

\begin{center}
$N=7$\\[2mm]\begin{tabular}{|>{\columncolor[gray]{0.8}}c|c|>{\columncolor[gray]{0.8}}c|>{\columncolor[gray]{0.8}}c|c|c|c|}
\hline
$c_{1111111}^{(7)}$&$c_{111112}^{(7)}$&$c_{11113}^{(7)}$&$c_{11122}^{(7)}$&$c_{1222}^{(7)}$&$c_{1123}^{(7)}$&$c_{1114}^{(7)}$\\\hline\hline
1&42&280&420&840&3360&1680\\ \hline
\end{tabular}\\[2mm]
\begin{tabular}{|>{\columncolor[gray]{0.8}}c|>{\columncolor[gray]{0.8}}c|>{\columncolor[gray]{0.8}}c|>{\columncolor[gray]{0.8}}c|c|c|c|>{\columncolor[gray]{0.8}}c|}\hline
$c_{223}^{(7)}$&$c_{133}^{(7)}$&$c_{115}^{(7)}$&$c_{124}^{(7)}$&$c_{34}^{(7)}$&$c_{25}^{(7)}$&$c_{16}^{(7)}$&$c_7^{(7)}$\\ \hline \hline
3360&4480&8064&10080&13440&16128&26880&46080\\ \hline
\end{tabular}
\end{center}

\begin{center}
$N=8$\\[2mm]
\begin{tabular}{|>{\columncolor[gray]{0.8}}c|c|>{\columncolor[gray]{0.8}}c|>{\columncolor[gray]{0.8}}c|c|c|c|}\hline
$c_{11111111}^{(8)}$ & $c_{1111112}^{(8)}$ & $c_{111113}^{(8)}$ & $c_{111122}^{(8)}$ &$c_{11222}^{(8)}$& $c_{11114}^{(8)}$&$c_{11123}^{(8)}$\\ \hline \hline
1&56&448&840&3360&3360&8960\\ \hline 
\end{tabular}\\[2mm]
\begin{tabular}{|>{\columncolor[gray]{0.8}}c|>{\columncolor[gray]{0.8}}c|>{\columncolor[gray]{0.8}}c|>{\columncolor[gray]{0.8}}c|>{\columncolor[gray]{0.8}}c|c|c|c|c|c|}\hline
$c_{2222}^{(8)}$&  $c_{1133}^{(8)}$ & $c_{1115}^{(8)}$&$c_{1223}^{(8)}$ & $c_{1124}^{(8)}$  &  $c_{116}^{(8)}$ & $c_{134}^{(8)}$& $c_{125}^{(8)}$  &$c_{224}^{(8)}$&$c_{233}^{(8)}$\\ \hline\hline
1680&17920&21504&26880&40320&107520&107520&129024&40320&35840\\\hline
\end{tabular}\\[2mm]
\begin{tabular}{|>{\columncolor[gray]{0.8}}c|>{\columncolor[gray]{0.8}}c|>{\columncolor[gray]{0.8}}c|>{\columncolor[gray]{0.8}}c|c|}\hline
$c_{44}^{(8)}$ & $c_{35}^{(8)}$ & $c_{26}^{(8)}$& $c_{17}^{(8)}$ & $c_{8}^{(8)}$ \\ \hline\hline
80640&172032&215040&368640&645120\\\hline
\end{tabular}
\end{center}

\begin{center}
$N=9$\\[2mm]
\begin{tabular}{|>{\columncolor[gray]{0.8}}c|c|>{\columncolor[gray]{0.8}}c|>{\columncolor[gray]{0.8}}c|c|c|c|}\hline
$c_{111111111}^{(9)}$	&$c_{11111112}^{(9)}$	&$c_{1111113}^{(9)}$	&$c_{1111122}^{(9)}$	&$c_{111114}^{(9)}$	&$c_{111222}^{(9)}$	&$c_{111123}^{(9)}$	 \\ \hline\hline
1				& 72			& 672			&1512			&6048			&10080			&20160						\\ \hline
\end{tabular} \\[2mm]
\begin{tabular}{|>{\columncolor[gray]{0.8}}c|>{\columncolor[gray]{0.8}}c|>{\columncolor[gray]{0.8}}c|>{\columncolor[gray]{0.8}}c|>{\columncolor[gray]{0.8}}c|c|c|c|c|c|c|}\hline
$c_{12222}^{(9)}$	&$c_{11115}^{(9)}$	&$c_{11133}^{(9)}$&$c_{11124}^{(9)}$		& $c_{11223}^{(9)}$	&$c_{2223}^{(9)}$	&$c_{1233}^{(9)}$	&$c_{1116}^{(9)}$	&$c_{1224}^{(9)}$	&$c_{1134}^{(9)}$	&$c_{1125}^{(9)}$	 \\ \hline\hline
15120				&48384			&53760	&120960			&120960			&80640		&322560		&322560	&362880	&483840	&580608	\\ \hline 
\end{tabular}\\[2mm]
\begin{tabular}{|>{\columncolor[gray]{0.8}}c|>{\columncolor[gray]{0.8}}c|>{\columncolor[gray]{0.8}}c|>{\columncolor[gray]{0.8}}c|>{\columncolor[gray]{0.8}}c|>{\columncolor[gray]{0.8}}c|>{\columncolor[gray]{0.8}}c|}\hline
$c_{333}^{(9)}$ &$c_{225}^{(9)}$&$c_{144}^{(9)}$	&$c_{234}^{(9)}$	&$c_{135}^{(9)}$	&$c_{117}^{(9)}$	&$c_{126}^{(9)}$	\\\hline\hline
143360	&580608&725760		&967680		&1548288	&1658880	&1935360	\\\hline
\end{tabular}\\[2mm]
\begin{tabular}{|c|c|c|c|>{\columncolor[gray]{0.8}}c|}\hline
$c_{45}^{(9)}$	&$c_{36}^{(9)}$	&$c_{27}^{(9)}$	&$c_{18}^{(9)}$	&$c_{9}^{(9)}$\\\hline\hline
2322432	&2580480	&3317760	&5806080	&10321920\\ \hline
\end{tabular}
\end{center}


\begin{thebibliography}{99}
\bibitem{OEIS}
The Online Encyclopedia of Integer Sequences \href{https://oeis.org/}{https://oeis.org/}
\bibitem{Diestel}
Diestel, Reinhard. Graph Theory, Springer Berlin/Heidelberg, 2024. ProQuest \href{https://ebookcentral.proquest.com/lib/durham/detail.action?docID=6310518}{Ebook Central}.
%\cite{Gehrmann-DeRidder:2003pne}
\bibitem{Gehrmann-DeRidder:2003pne}
A.~Gehrmann-De Ridder, T.~Gehrmann and G.~Heinrich,
%``Four particle phase space integrals in massless QCD,''
Nucl. Phys. B \textbf{682} (2004), 265-288
doi:10.1016/j.nuclphysb.2004.01.023
[arXiv:hep-ph/0311276 [hep-ph]].
%173 citations counted in INSPIRE as of 06 Mar 2025
%\cite{Alonso:2019ptb}
\bibitem{Alonso:2019ptb}
R.~Alonso and A.~Urbano,
%``Amplitudes, resonances, and the ultraviolet completion of gravity,''
Phys. Rev. D \textbf{100} (2019) no.9, 095013
doi:10.1103/PhysRevD.100.095013
[arXiv:1906.11687 [hep-ph]].
%11 citations counted in INSPIRE as of 06 Mar 2025
%\cite{Eden:1966dnq}
\bibitem{Eden:1966dnq}
R.~J.~Eden, P.~V.~Landshoff, D.~I.~Olive and J.~C.~Polkinghorne,
%``The analytic S-matrix,''
Cambridge Univ. Press, 1966,
ISBN 978-0-521-04869-9
%35 citations counted in INSPIRE as of 10 Mar 2025
%\cite{Arkani-Hamed:2017jhn}
\bibitem{Arkani-Hamed:2017jhn}
N.~Arkani-Hamed, T.~C.~Huang and Y.~t.~Huang,
%``Scattering amplitudes for all masses and spins,''
JHEP \textbf{11} (2021), 070
doi:10.1007/JHEP11(2021)070
[arXiv:1709.04891 [hep-th]].
%480 citations counted in INSPIRE as of 10 Mar 2025
\end{thebibliography}
\end{document}